\documentclass[useAMS,usenatbib]{mn2e}
\usepackage{graphicx}
\usepackage{multirow}

\title[CTCV J1300-3052]{A Radial Velocity Study of CTCV J1300-3052}
\author[C. D. J. Savoury et al.]
{C. D. J. Savoury$^{1}$\thanks{E-mail: chris.savoury@sheffield.ac.uk}, S. P. Littlefair$^{1}$, 
T. R. Marsh$^{2}$, V. S. Dhillon$^{1}$, S.G. Parsons$^{2}$, \newauthor C.M. Copperwheat$^{2}$ and D. Steeghs$^{2}$\\
$^{1}$Dept of Physics and Astronomy, University of Sheffield, Sheffield, S3 7RH, UK\\ 
$^{2}$Dept of Physics, University of Warwick, Coventry, CV4 7AL, UK}
\begin{document}

\date{Submitted for publication in the Monthly Notices of the Royal Astronomical 
Society \today}

\pagerange{\pageref{firstpage}--\pageref{lastpage}} 
\pubyear{2011}

\maketitle

\label{firstpage}

\begin{abstract}

We present time-resolved spectroscopy of the eclipsing, short period cataclysmic variable CTCV J1300-3052. 
Using absorption features from the secondary star, we determine the radial velocity semi-amplitude of 
the secondary star to be $K_{2}$ = 378 $\pm$ 6 km s$^{-1}$, and its projected rotational velocity to be 
$v\sin i = 125\pm7$ km s$^{-1}$. Using these parameters and Monte Carlo techniques, we obtain masses of 
$M_{1}$ = 0.79 $\pm$ 0.05 $M_{\odot}$ 
for the white dwarf primary and $M_{2}$ = 0.198 $\pm$ 0.029 $M_{\odot}$ for the M-type secondary star. These 
parameters are found to be in excellent agreement with previous mass determinations found via photometric 
fitting techniques, supporting the accuracy and validity of photometric mass determinations in short period
CVs.
 
\end{abstract}


\section{Introduction}
Cataclysmic variable stars (CVs) are a class of interacting binary system undergoing mass transfer from 
a Roche-lobe filling secondary to a white dwarf primary, usually via a gas stream and accretion disc.
Through determinations of the masses and radii of the component stars in CVs, it is possible to test 
fundamental theories regarding their formation, origin and evolution \citep[e.g.][]{littlefair2008, 
savoury2011}.

\citet{savoury2011} carried out a photometric study of eclipsing CVs and found the masses and radii for 
both the white dwarf and donor star in 14 systems. These masses were found by fitting a parameterised
model to the eclipse light curves. This model is based on the techniques developed by \citet{bailey1979}, 
\citet{smak1979}, \citet{cook1984}, \citet{wood1985}, \citet{wood1986}, \citet{horne1994}, \citet{littlefair2008} 
and \citet{copperwheat2010} and relies on just four assumptions: the bright-spot lies on the ballistic 
trajectory from the donor star; the donor star fills its Roche lobe; the white dwarf is accurately 
described by a theoretical mass-radius relation; and the whole of the white dwarf is visible with an unmodified 
surface brightness. It is photometric mass and radii determinations such as 
these that are used to calibrate the $\epsilon$-$q$ (super-hump excess-mass ratio) relations of \citet{patterson2005}, 
\citet{knigge2006} and \citet{knigge2011}, which can then be used to derive donor mass estimates for large samples of 
CVs. It is therefore important to check the validity of photometric mass determinations. 

For objects with periods above the period gap (the dearth of systems between 2.2 and 3.2 hours, see e.g. Kolb 
\& Ritter 2003, Knigge 2006)\nocite{ritter2003}, the photometric fitting technique appears robust, with donor 
star radial velocities predicted by photometry in agreement with those found by other techniques \citep{watson2003, 
feline2005, copperwheat2010}. However, for objects below the period gap, independent tests of the photometric 
technique are rare.\citet{tulloch2009} found the radial velocity of the white dwarf ($K_{1}$) in SDSS J143317.78+101123.3 
($P_{orb}$ = 78.1 mins), as measured from disc emission lines, to be in excellent agreement with the photometric 
value predicted by \citet{littlefair2008}. The agreement is encouraging, but the motion of the inner disc does 
not necessarily follow the motion of the white dwarf, and so $K_{1}$ estimates from disc emission should be treated 
with caution \citep[e.g.][]{marsh1988b}. More recently, \citet{copperwheat2011} found the radial velocity and rotational 
broadening of the secondary star in OY Car ($P_{orb}$ = 90.9 mins) to be in good agreement with those predicted by 
photometric methods \citep{wood1990, littlefair2008}. However, such is the importance of mass determinations in CVs, 
additional verification across a range of orbital periods is highly desirable.

One of the systems observed by \citet{savoury2011} was CTCV J1300-3052 (hereafter CTCV 1300). CTCV 1300 is a dwarf nova 
that was discovered as part of the Cal\'an-Tololo Survey follow up \citep{tappert2004}. It was found to be eclipsing, 
with an orbital period of 128.1 minutes, placing it immediately below the period gap. The average spectrum showed clear 
emission lines from the accretion disc and absorption lines from the donor star. It is through absorption lines such as 
these that we can determine the radial velocity and rotational broadening of the secondary star, which can in turn be 
used to derive an independent measure of the masses and radii of the component stars \citep[e.g.][]{horne1993, smith1998, 
thoroughgood2001, thoroughgood2004}.

In this paper we present time resolved-spectroscopy of CTCV 1300 and determine the system 
parameters. The parameters derived using spectroscopy will provide an independent test of the photometric
methods used by \citet{littlefair2008} and \citet{savoury2011}.

\section{Observations}
\label{sec:obs}

CTCV 1300 was observed using X-shooter \citep{odorico2006} in service mode mounted on UT2 (Kueyen) on the 8.2-m Very Large 
Telescope (VLT) on the nights beginning 9 Feb 2010 and 6 March 2010. In total, we obtained 48 spectra (24 on each night) 
covering 1.5 orbital cycles, and a wavelength range of $\sim$3000-24800 \AA. Exposure times were 235 seconds in the UVB-arm 
(3000-5500 \AA), 210 seconds in the VIS-arm (5500-10000 \AA), and 255 seconds in the NIR-arm (10000-24800 \AA), with dead 
times between exposures of approximately 8, 9 and 1 seconds, respectively. 

The target was observed with the 1.0''x11'' slit in the UVB-arm, the 1.2''x11'' slit in the VIS-arm, and the 0.9''x11'' 
slit in the NIR-arm. The resolving power was $\sim$5100 (59 km s$^{-1}$) in the UVB and NIR-arms, and $\sim$6700 (45 
km s$^{-1}$) in the VIS-arm. Seeing conditions on both nights were fair, varying between 0.5 and 1.5 arcseconds, but 
with flares of up to 2.0 arcseconds.

Observations of the standard star GD153 were used to flux calibrate the data and correct for telluric absorption. The 
data were obtained in `stare' mode rather than nodding along the slit as is normal for long slit infra-red spectroscopy. 
Consequently, the sky subtraction on the NIR-arm spectra is significantly worse than usual with X-shooter. We also obtained 
spectra of the spectral type templates GJ2066 (M2V) and GJ1156 (M5V), although these data were taken on the nights of 11 Dec 
2009 and 28 Jan 2010, respectively.

\section{Data Reduction \& Analysis} 

Data reduction was carried out using the X-shooter pipeline (version 1.2.2) recipes within ESORex, the ESO reduction
execution tool. The data for all three arms were reduced with similar procedures. The required calibration frames were 
constructed using the standard recipes provided in the pipeline. In brief, they include a map of bad pixels, a master 
bias (for the UVB and VIS-arms), a master dark (for the NIR-arm, as dark contribution is negligible in UVB and VIS) and 
a master flat. The data was first bias and dark subtracted, before an inter-order background was fitted and subtracted. 
Science frames were then divided by the flat field, and then the object was localised on the slit. Sky subtraction 
and cosmic ray removal took place, and the data then underwent an optimal extraction routine \citep{horne1986a, marsh1989}, 
and order merging. 

Wavelength calibration was undertaken using an arc line spectrum which was taken from the ESO archive. The wavelength calibration
recipe used a physical model to generate a best-guess solution of the line positions on the calibration frame. These lines were 
then fitted by 2D Gaussians, and the resulting positions of these lines were adjusted via a polynomial fit to the whole CCD.
From residuals to the line fitting, we estimate that this calibration is accurate to $\sim$1 km s$^{-1}$ (at $\lambda$ = 8183~\AA). 
We corrected for flexure in the VIS arm by measuring the shift of the observed sky lines relative to their positions measured by
\citet{hanuschik2003}. 
The individual spectra were then moved by these shifts, which were typically between 10 and 35 km s$^{-1}$. We do not attempt to 
correct for flexure in the UVB and NIR arms, since we do not attempt to measure radial velocities to a high degree of precision 
in these bands.

The time and wavelength axis of the data were corrected to the heliocentre. 

\section{Results}

\subsection{Average Spectra}
\begin{figure*}
\centering
\includegraphics[scale=0.60,angle=0,trim=0 0 0 0,clip]{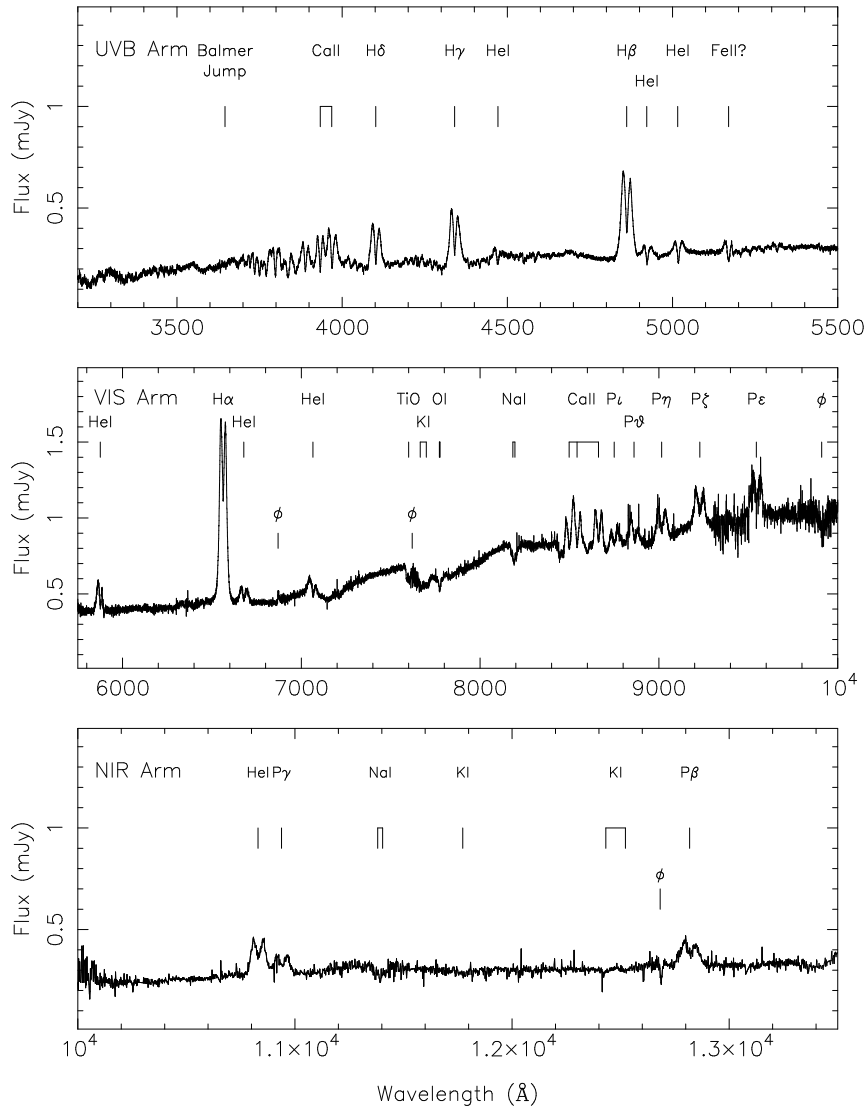}
\caption{The average spectra of CTCV 1300, in the rest frame of the binary. The upper panel shows the UVB-arm, the centre 
panel the VIS-arm, and the lower panel the NIR-arm. The most prominent features are labelled.}
\label{fig:ave_spectra}
\end{figure*}

The average spectra of CTCV 1300 are shown in Fig. \ref{fig:ave_spectra}. The upper panel shows the wavelength range 3200-5500
\AA~(from the UVB-arm), the centre panel shows 5750-10000 \AA~(from the VIS-arm), and the lower panel 10000-13500 \AA~(from the 
NIR-arm). Each spectrum is flux calibrated, and telluric correction has been attempted.

Throughout the spectrum we see strong, broad, double-peaked Balmer lines and several double-peaked He I lines 
(4471, 4922, 5015, 5875, 6678, 7065 and 10830 \AA). Broad double-peaked lines such as these are typical of a high-inclination 
accreting binary \citep[e.g.][]{horne1986}. 

\begin{figure*}
\centering
\includegraphics[scale=0.40,angle=-90,trim=0 0 0 0,clip]{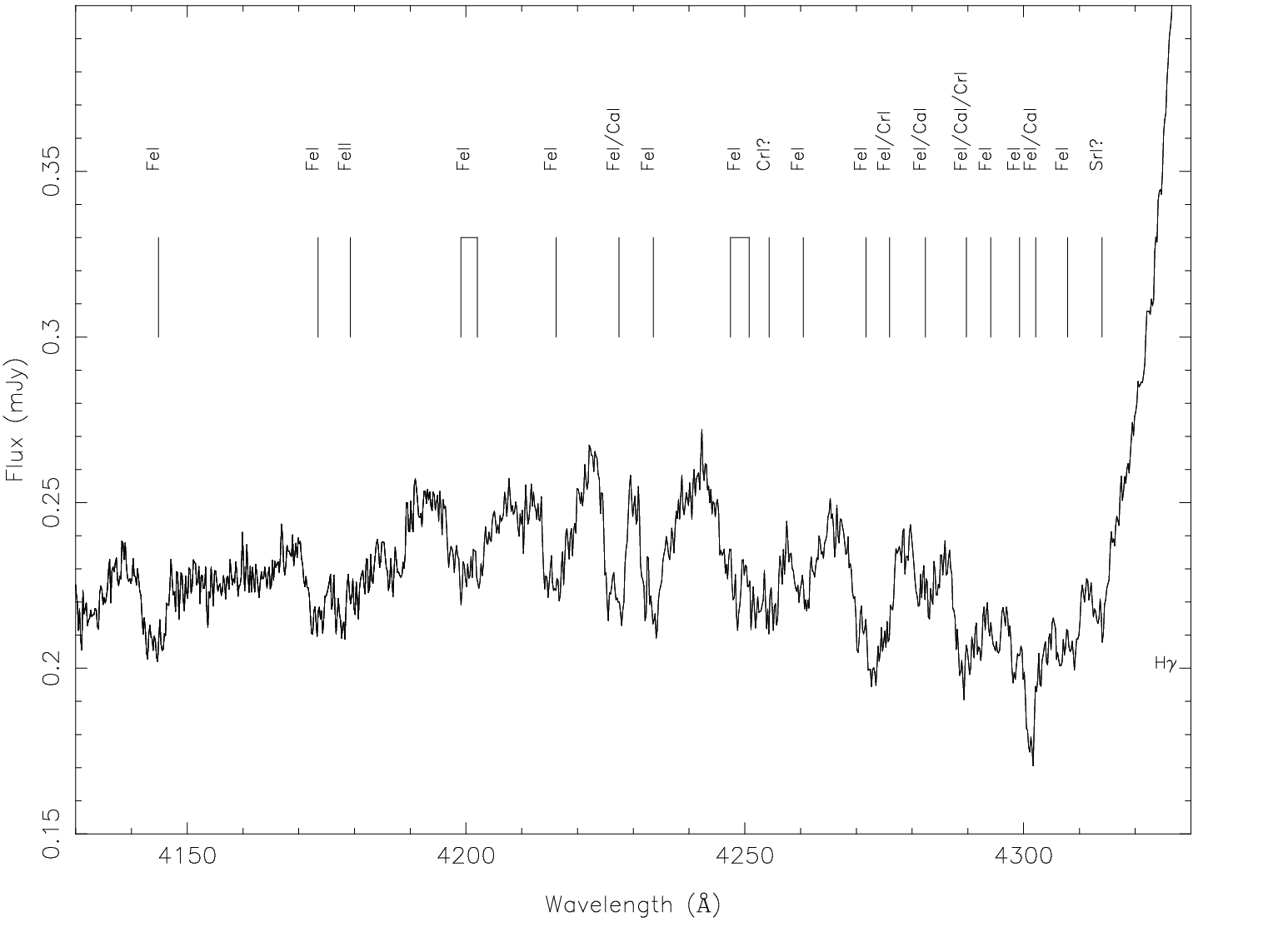}
\caption{The average spectra of CTCV 1300 between 4130-4330~\AA, corrected to the rest frame of the
white dwarf. Spectra taken during eclipse are not included in the average.}
\label{fig:wd_lines}
\end{figure*}

The high ionisation line He II 4686 \AA~appears absent in the average spectrum, but is visible in the trailed spectra 
(see Sections \ref{sec:trails} and \ref{sec:doppler}). Several absorption lines are present between 4000-4800 \AA~(see Fig. 
\ref{fig:wd_lines}) which appear to trace the motion of the disc (see Section \ref{sec:trails}). We believe the most likely cause 
of these absorption lines is a veil of disc material along the line of sight, The majority of these lines appear to be Fe I, 
Fe II and Ca I. Similar features have been observed in the spectrum of OY Car by \citet{horne1994} and \citet{copperwheat2011}.

The helium lines at 4922, 5015~\AA~appear to show strong, narrow absorption cores that dip below the continuum, as do the higher-order 
Balmer lines between 3600-4000 \AA. The O I triplet at 7773 \AA~is clearly visible, and also appears to drop below the 
continuum. Features such as these are observed in a number of CVs \citep[e.g][]{marsh1987, wade1988, friend1988}. These absorption cores 
are believed to originate through self-absorption in the accretion disc.

The Ca II triplet at 8498, 8542 and 8662 \AA~(hereafter 8567~\AA) is clearly present and originates from the disc, although there 
is evidence of emission from the irradiated side of the donor (see Sections \ref{sec:trails} and \ref{sec:doppler}). 
Similar features have been observed
in the spectrum of GW Lib \citep{spaandonk2010}. The higher orders of the Paschen 
series are also visible from $\sim$8800 \AA~onwards, and are possibly blended with the Ca II emission. 

Absorption features from the secondary star are clearly visible in the form of TiO bands around 7100~\AA~and 7600~\AA, and weak K I 
absorption doublet at 7664, 7699 (hereafter 7682~\AA), 11773, 12432 and 12522~\AA. However, these regions are heavily affected by telluric 
absorption. The clearest features from the secondary star are the Na I doublets at 8183, 8194~\AA~(hereafter 8189~\AA) and 
11381, 11404~\AA~(hereafter 11393~\AA), although the second of these is also heavily affected by telluric absorption.

\subsection{Trailed Spectra}
\label{sec:trails}

The data were phase binned into 30 bins, according to the ephemeris of \citet{savoury2011}. The UVB-arm has 
complete phase coverage although, due to the differing exposure times (see Section \ref{sec:obs}), one of these
bins is empty in both the VIS and NIR-arms.

\begin{figure*}
\centering
\includegraphics[scale=0.75,angle=0,trim=0 0 0 0,clip]{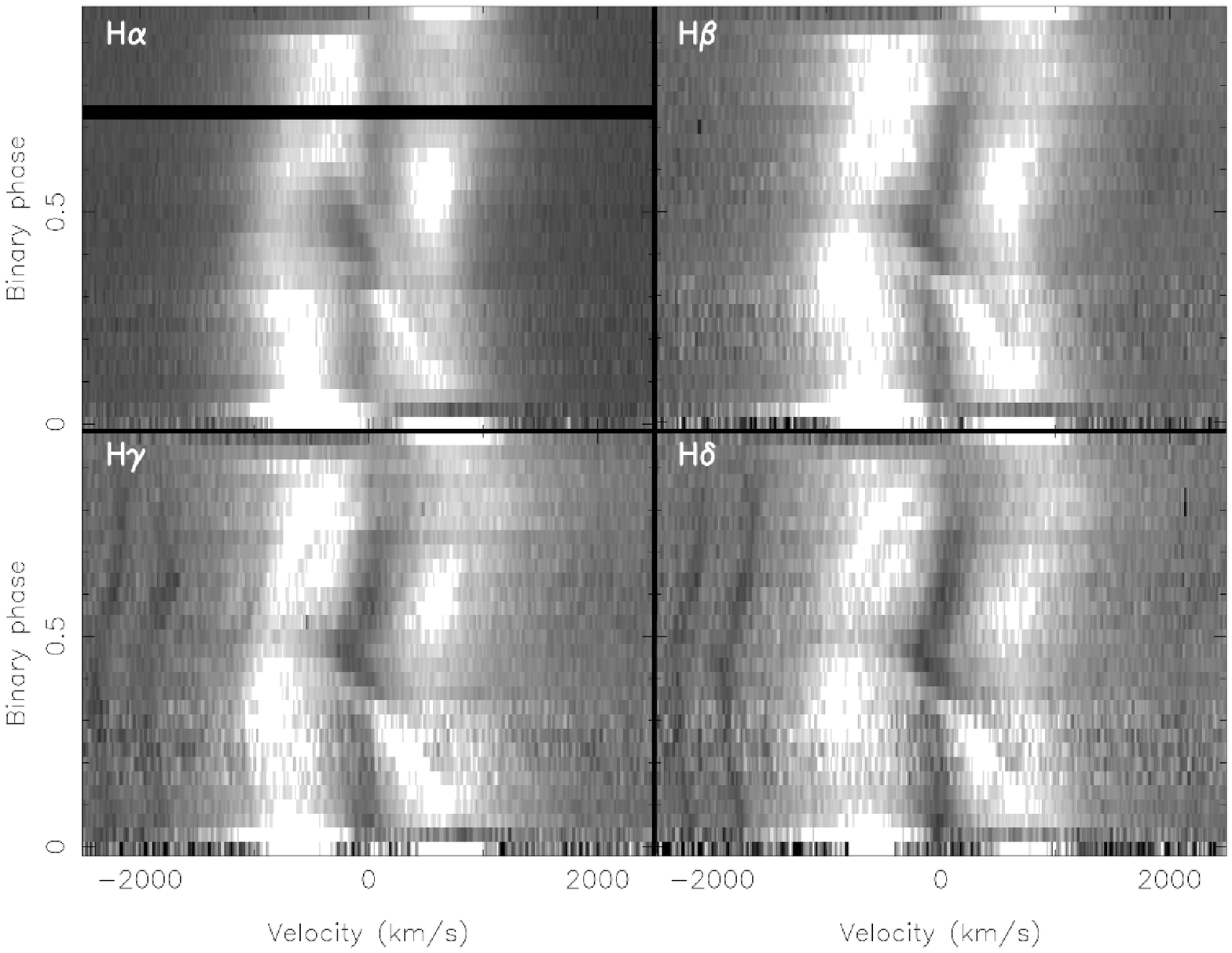}
\caption{The trailed spectra of the H$\alpha$ (top left), H$\beta$ (top right), H$\gamma$ (bottom left) and H$\delta$ 
(bottom right) lines in CTCV 1300.}
\label{fig:hydrogen_trails}
\end{figure*}

We divided the continuum by a polynomial and re-binned the spectra onto a constant velocity-interval scale centred on 
the rest wavelength of the lines. Fig. \ref{fig:hydrogen_trails} shows the trailed spectra of the H$\alpha$, H$\beta$, 
H$\gamma$ and H$\delta$ lines in CTCV 1300. Each line shows two clear peaks that vary sinusoidally with phase, in
addition to the characteristic s-wave between phases 0.1--0.4 from the bright spot.

\begin{figure*}
\centering
\includegraphics[scale=0.60,angle=0,trim=0 0 0 0,clip]{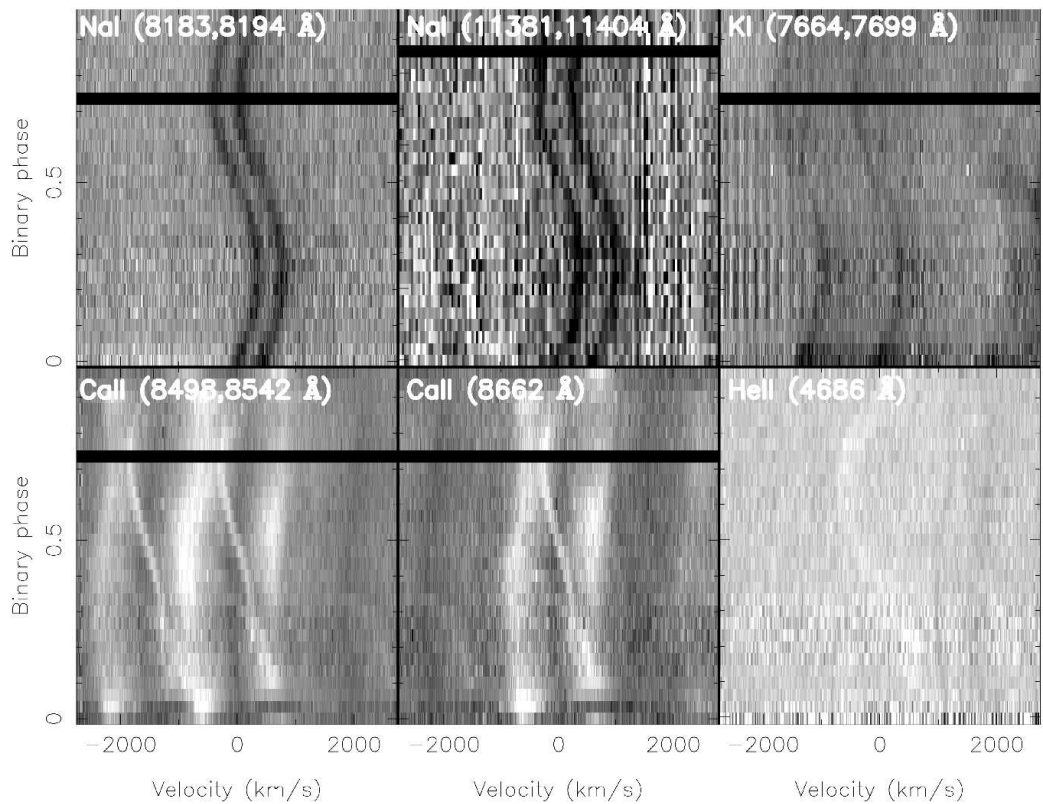}
\caption{The trailed spectra of the 8189 and 11393~\AA~Na I doublets (upper left and upper centre, respectively), the 7682~\AA~K I 
doublet (top right), the Ca II triplet (8498, 8542~\AA~bottom left, 8662~\AA~bottom centre) and He II (4686~\AA, bottom right) in CTCV 1300. 
Black and white lines represent absorption and emission, respectively.}
\label{fig:donor_trails}
\end{figure*}

In Fig. \ref{fig:donor_trails} we show the trailed spectra of two Na I doublets (8189 and 11393~\AA), the Ca II triplet (8567~\AA), 
the 7682~\AA~K I doublet and He II (4686~\AA). The phases at which the Na I and K I lines show maximum red-shift ($\phi = 0.25$) 
and blue-shift ($\phi = 0.75$) suggest they originate from the donor star. We see evidence for emission from the donor star in the 
Ca II lines through a component in the trail that is in phase with the Na I lines. However, this component is only visible 
during phases $\sim$0.25-0.75, which indicates it arises from the irradiated side of the donor. The He II line appears to follow the 
motion of the bright spot, as defined by the s-wave in the Balmer trails.

\begin{figure}
\centering
\includegraphics[scale=0.35,angle=-90,trim=0 0 0 0,clip]{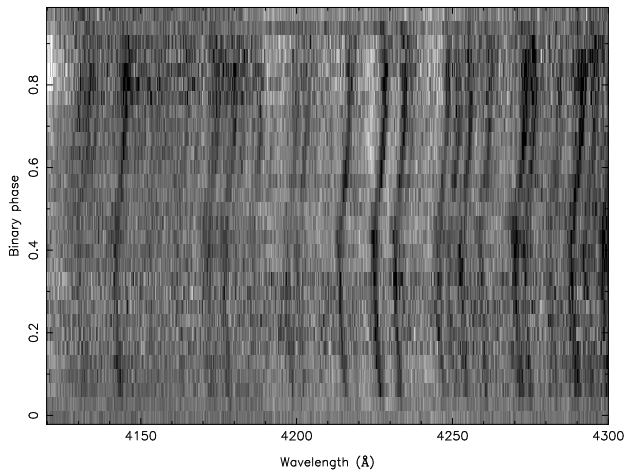}
\caption{Trailed spectra of the forest of FeI, FeII and CaI absorption lines between 4130-4300~\AA.}
\label{fig:abs_forest_trail}
\end{figure}

In Fig. \ref{fig:abs_forest_trail} we show the trailed spectra of the absorption line forest between 4130-4300~\AA. The lines
all appear to move together, suggesting a common place of origin. Using the same method outlined in Section \ref{sec:kr},
we find the velocity of these lines to be $K_{abs}$ = 116 $\pm$ 4 km s$^{-1}$, with a phase offset of 
$\Delta$$\phi$ = 0.072 $\pm$ 0.006. The high velocity (compared to the expected motion of the white dwarf, $\sim$ 90 km s$^{-1}$, see Section 
\ref{sec:system_params}) and significant phase offset suggests that these lines originate in the disc.

\subsection{Doppler tomography}
\label{sec:doppler}

Doppler tomography is an indirect imaging technique which can be used to determine the velocity-space distribution
of the emission in cataclysmic variables. For a comprehensive review of Doppler tomography, see \citet{marsh1988} 
and \citet{marsh2001}.

\begin{figure*}
\centering
\includegraphics[scale=0.75,angle=0,trim=0 0 0 0,clip]{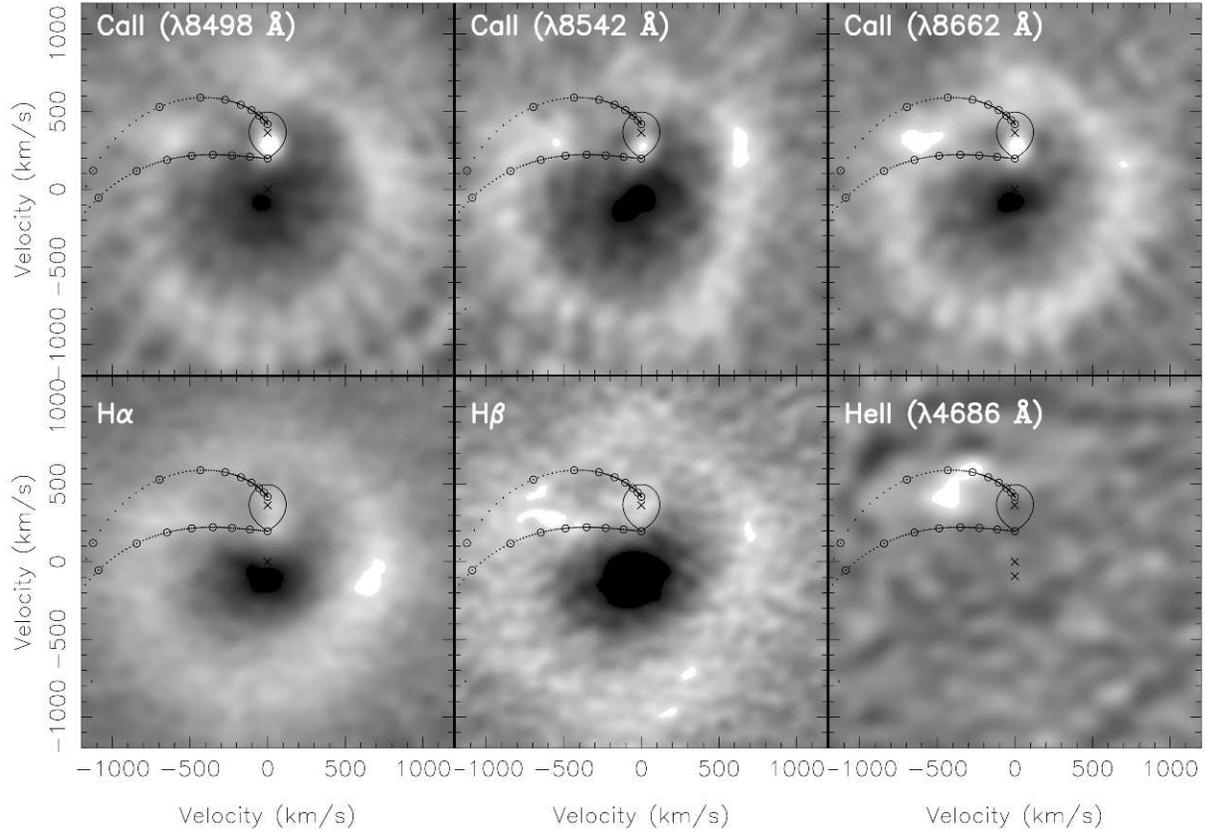}
\caption{Doppler maps of CTCV 1300 in Ca II (8498, 8542, 8662~\AA), H$\alpha$, H$\beta$ and He II (4686~\AA) computed 
from the trailed spectra in Figs. \ref{fig:hydrogen_trails} and \ref{fig:donor_trails}. 
Data taken during eclipse have been ommited from the fit. The predicted position of the secondary star and the path of 
the gas stream are marked. The three crosses on the map are, from top to bottom, the centre of mass of the secondary 
star, the centre ofmass of the system, and the white dwarf. These crosses, the Roche lobe of the secondary, the 
Keplerian velocity along the gas stream (top curve), and the predicted trajectory of the gas-stream (bottom curve) have 
been plotted using the system parameters found in Section \ref{sec:system_params}. The series of circles along the gas 
stream mark the distance from the white dwarf at intervals of 0.1$L_{1}$, where 1.0$L_{1}$ is the secondary star.}
\label{fig:doppler_maps}
\end{figure*}

Fig. \ref{fig:doppler_maps} shows Doppler maps for Ca II (8498, 8542 \& 8662~\AA), H$\alpha$, H$\beta$ and
He II (4686~\AA). Eclipse data (between phases 0.95 and 1.05) are removed. A systemic velocity of $\gamma$ = -20 km s$^{-1}$
was applied to shift the maps onto the $K_{x} = 0$ km s$^{-1}$ axis (see Section \ref{sec:kr}).

In each map we see a ring-like distribution of emission centred on the white dwarf, which is characteristic of an 
accretion disc. In the Ca II maps, we see an enhanced area of emission at velocities intermediate to the free-fall velocity 
of the gas stream (lower stream) and the velocity of the disc along the gas stream (Keplerian velocity, upper stream). This emission is attributed 
to the bright spot. The three Ca II maps all show clear emission from the donor star, which appears to be concentrated 
towards the inner hemisphere, indicating that irradiation is significant. The H$\alpha$ map shows emission from the secondary 
star, a feature uncommon in short period CVs. The H$\beta$ map shows weak bright spot emission.
The He II emission appears to show emission near both the Keplerian velocity stream and at velocities intermediate to the Keplerian 
velocity stream and free-fall velocity stream, although it is possible that this is an artifact arising from limited phase coverage 
\citep{marsh1988}.
If this is a genuine feature, its position relative to the Ca II emission suggests that the He II emission is caused 
by a mixture of gas-stream and disc material. He II bright spot emission at Keplerian disc velocities has been observed in 
other short period CVs \citep{marsh1990, copperwheat2011}. 

\subsection{Radial velocity of the secondary star}
\label{sec:kr}

The secondary star in CTCV 1300 is visible through weak absorption lines. The strongest of these lines is the Na I doublet 
at 8189~\AA. In order to determine the radial velocity, we cross-correlated the individual spectra of CTCV 1300 against an 
average spectra of CTCV 1300 using an iterative technique. We chose the Na I line at 8189~\AA~because it is much 
stronger and less affected by telluric absorption than the line at 11393~\AA.

\begin{figure}
\centering
\includegraphics[scale=0.4,angle=0,trim=0 0 0 0,clip]{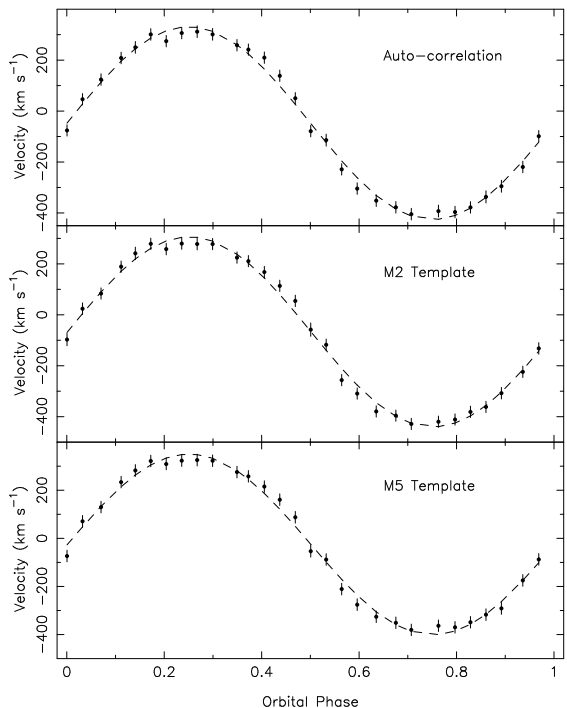}
\caption{The radial velocity curve of CTCV 1300 obtained through auto-correlation (upper panel), 
cross-correlation against an M2 template (centre panel) and cross-correlation against an M5 template (lower panel).}
\label{fig:kr}
\end{figure}

We subtracted fits to the continuum from the individual spectra and then corrected for the orbital motion of the secondary 
star with a first guess of $K_{2}$. For each individual spectrum, we then created a template spectrum that consisted of an 
average of all the spectra {\it minus} the spectrum under study \citep{marsh1994}. These template spectra were then cross-correlated 
against the uncorrected data. The velocity shifts as a function of orbital phase were then fit with a sine function according to;
\begin{equation}
\label{equation:gamma}
  V = \gamma - K_{2}\sin [2\pi(\phi - \phi_{0})],
\end{equation}
where V is the velocity shift, $\gamma$ is the systemic velocity of the system, $K_{2}$ is the radial velocity of secondary star, 
$\phi$ is the orbital phase, and $\phi_{0}$ is the phase offset. This then yielded a new value of $K_{2}$ and $\gamma$ to correct our 
spectra with. We added an intrinsic error in quadrature to each error bar to account for systematic error, and reach a reduced-$\chi^{2}$ of 
1. This process was repeated until $K_{2}$ converged. We arrive at a value of $K_{2}$ = 379 $\pm$ 6 km s$^{-1}$, with an intrinsic 
error of 22 km s$^{-1}$ added in quadrature to each error bar. The radial velocity curve obtained using this technique is shown in 
the upper panel of Fig. \ref{fig:kr}. 

The value of $\gamma$ obtained via our auto-correlation technique is not representative of the true systemic velocity, which must be
determined through cross-correlation with a template star of known radial velocity. Therefore, in order to verify this result and find 
the true systemic velocity ($\gamma$), we then cross-correlated against our M-dwarf 
template spectra using the same wavelength range. The template spectra were artificially broadened by 46 km s$^{-1}$ to account 
for the orbital smearing of CTCV 1300 through the 210-second VIS-arm exposures, and then by the best-fitting values for the 
rotational velocity of the secondary star ($v\sin i$) found in Section \ref{sec:donor_rotation}. An intrinsic error of 22 
km s$^{-1}$ was added to each error bar from the M2V cross-correlation, and 24 km s$^{-1}$ to M5V data, to account for systematic 
errors and reach a reduced-$\chi^{2}$ of 1. The radial velocity curves are shown in the centre panel (M2), and bottom panel (M5) 
of Fig. \ref{fig:kr}. Cross-correlating against the M2 and M5 templates yield values of $K_{2}$ = 373 $\pm$ 6 km s$^{-1}$ 
and $K_{2}$ = 376 $\pm$ 7 km s$^{-1}$, respectively. The M2 template could not be corrected for flexure, so we only use the 
M5 template to derive the systemic velocity for CTCV 1300. Using the radial velocities provided by \citet{gizis2002}, we find
$\gamma$ = -20 $\pm$ 5 km s$^{-1}$. For $K_{2}$, we prefer the value found through the auto-correlation, that is $K_{2}$ = 379 $\pm$  
6 km s$^{-1}$. This is because the average spectra of the data is a better match to the data than the M5 and M2 templates. 

The radial velocity curves produced through this technique show some variation from a sine fit between phases 0.4 
to 0.6, which is characteristic of irradiation suppressed absorption \citep[e.g.][]{billington1996}. \citet{marsh1988}
recommend only fitting the above data between phases 0.8 to 1.2, since at these phases the effects of irradiation are at a minimum.
Fitting the auto-correlation data, we obtain a value of $K_{2}$ = 378 $\pm$ 6 km s$^{-1}$, which we use hereafter. This 
value is consistent with the value predicted by \citet{savoury2011}, $K_{2}$ = 372.2 $\pm$ 2.5 km s$^{-1}$. 
Fitting the M5 and M2 templates between the same phases gives $K_{2}$ = 372 $\pm$ 7 km s$^{-1}$ and $K_{2}$ = 378 $\pm$ 8 km s$^{-1}$
respectively, which is again in excellent agreement with the auto-correlation data, and consistent with the photometric method.

\subsection{Rotational velocity of the secondary star}
\label{sec:donor_rotation}

The normalised spectra of CTCV 1300 were corrected for the orbital motion of the secondary star using the value of 
$K_{2}$ obtained in Section \ref{sec:kr}. The spectra were then averaged together in order to maximise the strength 
of the Na I doublet at 8189~\AA. The spectral-type templates were broadened to match the smearing due to orbital 
motion of CTCV 1300 through the 210 second VIS-arm exposures and rotationally broadened by a range of velocities 
(50-200 km s$^{-1}$). In principle, the orbital smearing is a function of orbital phase, and thus varies throughout 
the orbital cycle. We use a single value of 46 km s$^{-1}$, which is the average value of the smearing across an orbital 
cycle. We find that changing this to the maximum and minimum possible values of orbital smearing required, that is the 
smearing at conjunction and quadrature, alters the final value of $v\sin i$ obtained by 3 km s$^{-1}$. This uncertainty 
is added in quadrature to the uncertainty calculated below. 

The value of $v\sin i$ was obtained via an optimal subtraction routine, which subtracts a constant times the normalised, 
broadened template spectrum from the normalised, orbitally corrected CV spectrum.  This constant is adjusted to minimise 
the residual scatter between the spectra. The scatter is measured by carrying out the subtraction and then computing 
$\chi^{2}$ between the residual spectrum and a smoothed version of itself. By finding the value of rotational broadening 
that minimises $\chi^{2}$, we can obtain a value of $v\sin i$ and the spectral type of the secondary star \citep{dhillon1993, 
marsh1994}. This value of $v\sin i$ should then be corrected for the intrinsic rotational velocity of the template star. 
Unfortunately a wide range of spectral-types were not available, and so we are unable to deduce the spectral-type 
of the secondary using this technique.

\begin{figure}
\centering
\includegraphics[scale=0.40,angle=0,trim=0 0 0 0,clip]{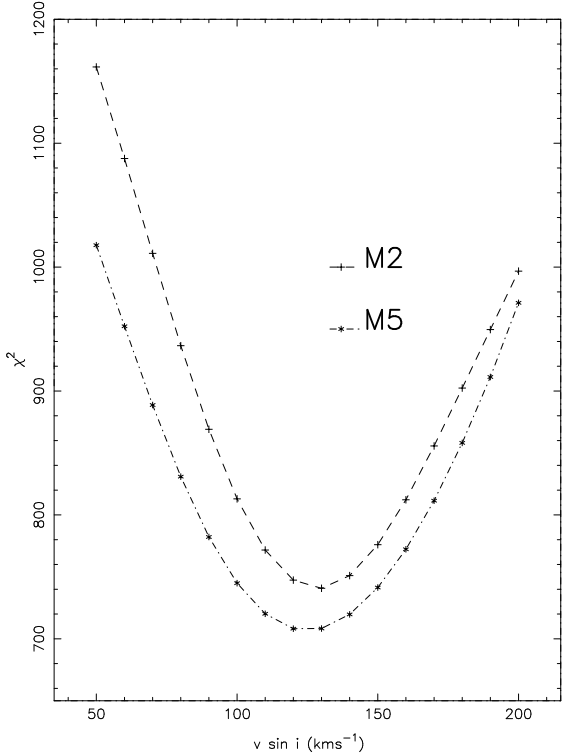}
\caption{$\chi^{2}$ vs $v\sin i$ from the optimal subtraction technique.}
\label{fig:optsub}
\end{figure}

The value of $v\sin i$ obtained using this method was found to vary depending on the spectral-type template used and 
the wavelength region selected for optimal subtraction. We attempted to include as much of the continuum as possible 
around the Na I doublet, while trying to avoid telluric regions. We used a wavelength range of 8080-8106, 8125-8206, 
8226-8245 and 8264-8285~\AA, a limb-darkening coefficient of 0.5 and smoothing Gaussian of FWHM = 15 km s$^{-1}$, which 
were found to give the lowest values of $\chi^{2}$. The limb-darkening coefficient is highly uncertain, although 
\citet{copperwheat2011} have shown that altering the limb-darkening coefficient has little effect on the value of $v\sin i$ 
obtained. We plot the values of $\chi^{2}$ versus $v\sin i$ for both spectral-type templates in Fig. \ref{fig:optsub}.
Using the M2 template, we obtain a value of $v\sin i$ = 129 $\pm$ 3 km s$^{-1}$, while the M5 template yields a value 
of $v\sin i$ = 125 $\pm$ 4 km s$^{-1}$. The uncertainties on these values come from the formal error estimation of  
$\Delta$$\chi^{2}$$=\pm1$. This does not attempt to include systematic errors. Because of the lack of available templates, 
we estimated the spectral type of the secondary star using the empirical donor sequence of \citet{knigge2011}. For a system 
with an orbital period of 128.07 minutes, we expect a secondary with spectral type of M4.3. We adopt a spectral type of M4.5 
$\pm$ 0.5. We interpolate between the two values of $v\sin i$ above to arrive at a final value of $v\sin i$ = 125 $\pm$ 7 
km s$^{-1}$. This error takes into account both the uncertainty on finding our minimum $v\sin i$ for each template ($\pm$ 3-4 
km s$^{-1}$ for each template), the uncertainty from averaging the orbital smearing ($\pm$ 3 km s$^{-1}$), and the uncertainty 
in spectral type ($\pm$0.5 spectral types).

\subsection{System Parameters}
\label{sec:system_params}

Using the values of $K_{2}$ = 378 $\pm$ 6 km s$^{-1}$ and $v\sin i$ = 125 $\pm$ 7 km s$^{-1}$ found in Sections \ref{sec:kr}
and \ref{sec:donor_rotation} in conjunction with the orbital period and a measurement of the eclipse full width at half
depth ($\Delta$$\phi$$_{1/2}$), we can calculate accurate system parameters for CTCV 1300.

The best measurement of the orbital period, $P_{orb}$, comes from \citet{savoury2011}, who determine 
$P_{orb}$ = 0.088940717(1) days. \citet{savoury2011} also present six light curves of CTCV 1300, from which we determine 
$\Delta$$\phi$$_{1/2}$ = 0.0791(5).

We use a Monte Carlo approach similar to \citet{horne1993}, \citet{smith1998}, \citet{thoroughgood2001} and \citet{thoroughgood2004} 
to calculate the system parameters and their errors. For a given set of $K_{2}$, $v\sin i$, $P_{orb}$ and $\Delta$$\phi$$_{1/2}$, the 
remaining parameters are calculated as follows.

$R_{2}/a$ can be estimated because the secondary star fills its Roche Lobe. $R_{2}$ is the secondary radius, 
and the $a$ is the binary separation, and so we use Eggleton's formula \citep{eggleton1983}, which gives the volume equivalent
radius of the Roche Lobe to an accuracy of $\sim$1 per cent, which is close to the equatorial radius of the secondary star as 
seen during eclipse,
\begin{equation}
\centering
\label{equation:r2a}
  \frac{R_{2}}{a} = \frac{0.49q^{2/3}}{0.6q^{2/3} + \ln(1+q^{1/3})}.
\end{equation}
The secondary star rotates synchronously with the orbital motion, so we can combine $K_{2}$ and $v\sin i$, to get
\begin{equation}
\centering
\label{equation:q}
  \frac{R_{2}}{a}(1+q) = \frac{v\sin i}{K_{2}}.
\end{equation}
This gives us two simultaneous equations that can be solved for $q$ and $R_{2}/a$. The orbital inclination, $i$, is fixed 
by $q$ and $\Delta$$\phi$$_{1/2}$, using geometrical arguements \citep[e.g.][]{bailey1979}. We determine the inclination
via a binary chop search using an accurate model of the Roche Lobe.

Using Kepler's Third Law, we obtain
\begin{equation}
\centering
\label{equation:keper}
  \frac{K_{2}^{3}P_{orb}}{2\pi G} = \frac{M_{1}\sin^{3}i}{(1+q)^{2}},
\end{equation}
which using the previously calculated values of $q$ and $i$ yields the mass of the primary star, $M_{1}$. The mass of the secondary star, $M_{2}$ 
and radial velocity of the primary, $K_{1}$, is given by
\begin{equation}
\centering
\label{equation:m2}
  q = \frac{M_{2}}{M_{1}} = \frac{K_{1}}{K_{2}}.
\end{equation}
Finally, we can calculate the radius of the secondary star using
\begin{equation}
\centering
\label{equation:r2}
  \frac{v\sin i}{R_{2}} = \frac{2\pi \sin i}{P_{orb}},
\end{equation}
and the binary separation, $a$, using equations \ref{equation:q} and \ref{equation:r2}.

\begin{figure*}
\centering
\includegraphics[angle=0,scale=0.50,trim=0 0 0 0,clip]{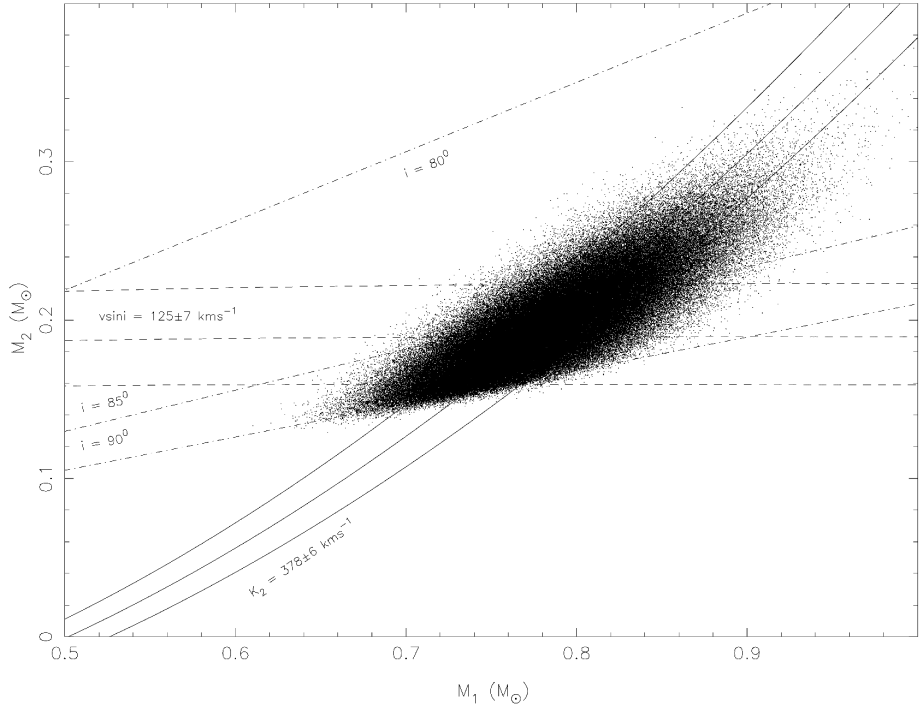}
\caption{Monte Carlo determination of system parameters for CTCV 1300. Each dot represents an ($M_{1}$, $M_{2}$) pair. 
Dot-dashed lines are lines of constant inclination, the solid curves satisfy the constraints from the radial velocity of the
secondary star, $K_{2}$, and the dashed lines satisfy the constraints of the rotational velocity of the secondary star, $v\sin i$.} 
\label{fig:monte_carlo}
\end{figure*}

Our Monte Carlo simulation takes 250,000 values of $K_{2}$, $\Delta\phi_{1/2}$, $v\sin i$ and $P_{orb}$, treating each as being normally
distributed about their measured values with standard deviations equal to the errors on the measurements. We then calculate the
mass of each component, the inclination of the system and the radius of the secondary star as outlined above,
omitting ($K_{2}$, $v\sin i$, $\Delta\phi_{1/2}$) triplets that are inconsistent with $\sin i$ $\le$1. Each accepted $M_{1}$, $M_{2}$ pair is  
plotted in Fig. \ref{fig:monte_carlo}, and the masses and their errors are computed from the mean and standard deviation of the distribution
of these pairs. We find that $M_{1}$ = 0.79 $\pm$ 0.05 $M_{\odot}$ and $M_{2}$ = 0.198 $\pm$ 0.029 $M_{\odot}$. These values are found to be in good
agreement with those of \citet{savoury2011}. The values of all system parameters found from the Monte Carlo simulation are listed in Table 
\ref{table:params}, along with those of \citet{savoury2011} for direct comparison.

\begin{table*}
\begin{center}
\caption{System parameters for CTCV 1300.} 
\begin{tabular}{lccc}
\hline
Parameter            & Measured Values & Monte Carlo Values & \citet{savoury2011} \\
\hline
$P_{orb}$ (s)         & - & - & $0.088940717(1)$\\
$\Delta\phi_{1/2}$    & $0.0791\pm0.0005$ & - & {\it not stated}\\
$K_{2}$ (km s$^{-1}$)   & $378\pm6$    & - & $372.2\pm2.5$\\
$v\sin i$ (km s$^{-1}$) & $125\pm7$       & - & $122\pm10$$^{*}$\\
\hline
$q$                  & - & $0.252\pm0.025$ & $0.240\pm0.021$\\
$i^{o}$              & - & $85.7\pm1.5$ &  $86.3\pm1.1$\\
$M_{1}/M_{\odot}$     & - & $0.79\pm0.05$ & $0.736\pm0.014$\\
$M_{2}/M_{\odot}$     & - & $0.198\pm0.029$ & $0.177\pm0.021$\\
$R_{2}/R_{\odot}$     & - & $0.223\pm0.011$ & $0.215\pm0.008$\\
$a/R_{\odot}$         & - & $0.834\pm0.020$ & $0.813\pm0.011$\\
$K_{1}$ (km s$^{-1}$ ) & - & $95\pm9$ & $90\pm8$ \\
Distance (pc)        & $330\pm40$ & - & $375\pm13$ \\
\hline
\end{tabular}
\label{table:params}
\end{center}
{\footnotesize $^{*}$ Derived using the values published in table 3 of \citet{savoury2011}.}
\end{table*}

\subsection{Distance}

By finding the apparent magnitude of the secondary star from its contribution to the total light during eclipse, and by estimating
the absolute magnitude, we can calculate the distance ($d$), using the equation;
\begin{equation}
\centering
  5\log (d/10) = m_{I} - M_{I} -dA_{I}/1000,
\end{equation}
where $A_{I}$ is the interstellar extinction in magnitudes per kpc. We assume the extinction is zero, as this allows 
a direct comparison to the distance obtained by \citet{savoury2011}, who used model generated white dwarf fluxes to
estimate the distance without correction for extinction. At mid-eclipse ($\phi = 0$), the apparent magnitude of the system 
is 17.32 $\pm$ 0.02 around the Na I doublet, which is approximately the $I$-band. This value is not corrected for slit losses. 
The secondary star is found to contribute 58 $\pm$ 6 per cent, which gives an apparent magnitude of $m_{I}$ = 17.91 $\pm$ 0.09. 
We estimate the absolute magnitude using the empirical donor sequence of \citet{knigge2011}, who assume the donor is on the 
main sequence and then correct for bloating effects. From this, we take $M_{I} = 10.32\pm0.14$, and obtain a distance of 
$d$ = 330 $\pm$ 40 pc. This distance is found to be in good agreement with that of \citet{savoury2011}, who obtained a value 
of $d$ = 375 $\pm$ 13 pc. 

\section{Discussion}
\label{sec:disc}

The system parameters listed in Table \ref{table:params} are found to be in good agreement with those of \citet{savoury2011}.
Together with \citet{copperwheat2011}, this gives us confidence that photometric mass determinations such as those of 
\citet{littlefair2008} and \citet{savoury2011} are reliable across a range of orbital periods, and that the $\epsilon$-$q$ 
relations of \citet{patterson2005}, \citet{knigge2006} and \citet{knigge2011} are well founded. 

The uncertainties in the system parameters for CTCV 1300 determined in this paper, and in \citet{savoury2011}, are quite large in 
comparison to many of the other systems published in \citet{savoury2011}. The reason for the large uncertainties in CTCV 1300 in 
\citet{savoury2011} is because the eclipse light curves used for model fitting suffer from heavy flickering, which causes 
difficulties in obtaining an accurate value for the mass ratio, $q$. 
The large uncertainties in this paper arise because of the interpolation technique used to arrive at a value for $v\sin i$. The 
error on $v\sin i$ ($\pm$ 7 km s$^{-1}$) is the dominant source of uncertainty in our final system parameters. In principle, a wider 
selection of spectral type templates would enable us to further constrain $v\sin i$, and derive the spectral type.

\section{Conclusions}

We have used time-resolved spectroscopy to determine the system parameters for the short period dwarf nova CTCV 
1300. The double-peaked nature of the Balmer and He I lines confirms the presence of an accretion disc, while careful
analysis of the Na I doublet absorption lines at 8189~\AA~reveals the radial velocity of the secondary star to be 
$K_{2}$ = 378 $\pm$ 6 km s$^{-1}$ and the rotational velocity of the secondary star to be $v\sin i$ = $125\pm7$ km s$^{-1}$. 
Using these measurements, we find $M_{1}$ = 0.79 $\pm$ 0.05 $M_{\odot}$ for the white dwarf primary and $M_{2}$ = 0.198 $\pm$ 
0.029 $M_{\odot}$ for the M-type secondary star. The radius of the secondary star is found to be $R_{2}$ = 0.223 $\pm$ 0.011 $R_{\odot}$. 

The system parameters determined through spectroscopic analysis are found to be in good agreement with those previously calculated
using photometric techniques. This is significant, as our results support the validity and accuracy of the purely photometric mass
determination technique in short period cataclysmic variables.

\section{Acknowledgements}
CDJS, VSD, TRM and CMC acknowledge the support of the Science and Technology Facilities Council (STFC). SPL acknowledges the 
support of an RCUK Fellowship. DS acknowledges a STFC Advanced Fellowship. This article is based upon observations carried out
using the European Southern Observatory (Paranal, Chile) with X-shooter on VLT-UT2 [programme 084.D-1149]. This research has 
made use of NASA's Astrophysics Data Bibliographic Services. 
\bibliographystyle{mn2e}
\bibliography{referencelist}

\begin{thebibliography}{}

\bibitem[\protect\citeauthoryear{Bailey}{Bailey}{1979}]{bailey1979}
Bailey J., 1979, MNRAS, 187, 645

\bibitem[\protect\citeauthoryear{Billington, Marsh \& Dhillon}{Billington et~al.}{1996}]{billington1996}
Billington I., Marsh T., Dhillon V., 1996, MNRAS, 278, 673

\bibitem[\protect\citeauthoryear{Cook \& Warner}{Cook \& Warner}{1984}]{cook1984}
Cook C., Warner B., 1984, MNRAS, 207, 705

\bibitem[\protect\citeauthoryear{Copperwheat, Marsh, Dhillon, Littlefair, Hickman, G{\"a}nsicke \& Southworth}{Copperwheat
  et~al.}{2010}]{copperwheat2010}
Copperwheat C.,  Marsh T.,  Dhillon V.,  Littlefair S.,  Hickman R.,
  G{\"a}nsicke B.,    Southworth J.,  2010, MNRAS, 402, 1824

\bibitem[\protect\citeauthoryear{Copperwheat et~al}{Copperwheat et~al.}{2011}]{copperwheat2011}
Copperwheat C. et~al., 2011, MNRAS, {\it submitted}

\bibitem[\protect\citeauthoryear{Dhillon, Marsh \& Jones}{Dhillon et~al.}{1991}]{dhillon1991}
Dhillon V.,  Marsh T.,    Jones D.,  1991, MNRAS, 252, 342

\bibitem[\protect\citeauthoryear{Dhillon \& Marsh}{Dhillon \& Marsh}{1993}]{dhillon1993}
Dhillon V.,  Marsh T., 1993, in Cataclysmic Variables and Related Physics, 2nd Technion Haifa Conference. 
Edited by O. Regev and Giora Shaviv. Annals of the Israel Physical Society, Volume 10. Bristol: 
Institute of Physics Pub.; Jerusalem: Israel Physical Society; New York: American Institute of Physics, 34

\bibitem[\protect\citeauthoryear{D'Odorico et~al}{D'Odorico et~al.}{2006}]{odorico2006}
D'Odorico S. et~al., 2006, in Society of Photo-Optical Instrumentation
  Engineers (SPIE) Conference Series Vol.~6269 of Society of Photo-Optical
  Instrumentation Engineers (SPIE) Conference Series, X-shooter uv- to k-band
  intermediate-resolution high-efficiency spectrograph for the vlt: status
  report at the final design review

\bibitem[\protect\citeauthoryear{Eggleton}{Eggleton}{1983}]{eggleton1983}
Eggleton P.,  1983, ApJ, 268, 368

\bibitem[\protect\citeauthoryear{Feline}{Feline}{2005}]{feline2005}
Feline W.,  2005, PhD Thesis, Univ. Sheffield

\bibitem[\protect\citeauthoryear{Friend et~al}{Friend et~al.}{1988}]{friend1988}
Friend, M., Martin J., Smith R., Jones D., 1988, MNRAS, 233, 451 

\bibitem[\protect\citeauthoryear{Gizis, Reid \& Hawley}{Gizis et~al.}{2002}]{gizis2002}
Gizis J., Reid N., Hawley S., 2002, AJ, 123, 3356 

\bibitem[\protect\citeauthoryear{Hanuschik}{Hanuschik}{2003}]{hanuschik2003}
Hanuschik R., 2003, A\&A, 407, 1157

\bibitem[\protect\citeauthoryear{Horne}{Horne}{1986}]{horne1986a}
Horne K., 1986, PASP, 98, 609

\bibitem[\protect\citeauthoryear{Horne \& Marsh}{Horne \&  Marsh}{1986}]{horne1986}
Horne K.,  Marsh T., 1986, MNRAS, 218, 761

\bibitem[\protect\citeauthoryear{Horne, Welsh \& Wade}{Horne et~al.}{1993}]{horne1993}
Horne K.,  Welsh W.,    Wade R.,  1993, ApJ, 410, 357

\bibitem[\protect\citeauthoryear{Horne, Marsh, Cheng, Hugeny \& Lanz}{Horne et~al.}{1994}]{horne1994}
Horne K.,  Marsh T.,  Cheng F.,  Hugeny I.,    Lanz T.,  1994, ApJ, 426, 294

\bibitem[\protect\citeauthoryear{Knigge}{Knigge}{2006}]{knigge2006}
Knigge C.,  2006, MNRAS, 373, 484

\bibitem[\protect\citeauthoryear{Knigge, Baraffe \& Patterson}{Knigge et~al.}{2011}]{knigge2011}
Knigge C.,  Baraffe I.,    Patterson J.,  2011, ApJS, 194, 28

\bibitem[\protect\citeauthoryear{Littlefair, Dhillon, Marsh, G{\"a}nsicke,
  Southworth, Baraffe, Watson \& Copperwheat}{Littlefair et~al.}{2008}]{littlefair2008}
Littlefair S.,  Dhillon V.,  Marsh T.,  G{\"a}nsicke B.,  Southworth J.,
  Baraffe I.,  Watson C.,    Copperwheat C.,  2008, MNRAS, 388, 1582

\bibitem[\protect\citeauthoryear{Marsh}{Marsh}{1987}]{marsh1987}
Marsh T.,  1987, MNRAS, 228, 779

\bibitem[\protect\citeauthoryear{Marsh}{Marsh}{1988}]{marsh1988b}
Marsh T., 1988, MNRAS, 231, 1117

\bibitem[\protect\citeauthoryear{Marsh}{Marsh}{1989}]{marsh1989}
Marsh T., 1989, PASP, 101, 1032

\bibitem[\protect\citeauthoryear{Marsh \& Horne}{Marsh \& Horne}{1988}]{marsh1988}
Marsh T., Horne K., 1988, MNRAS, 235, 269

\bibitem[\protect\citeauthoryear{Marsh}{Marsh}{2001}]{marsh2001}
Marsh T.,  2001, in {H.~M.~J.~Boffin, D.~Steeghs, \& J.~Cuypers} ed.,
  Astrotomography, Indirect Imaging Methods in Observational Astronomy Vol.~573
  of Lecture Notes in Physics, Berlin Springer Verlag, Doppler tomography.
pp~1--+

\bibitem[\protect\citeauthoryear{Marsh et~al}{Marsh et~al.}{1990}]{marsh1990}
Marsh T., Horne K., Schlegel E., Honeycutt R., Kaitchuck R., 1990, ApJ, 637

\bibitem[\protect\citeauthoryear{Marsh, Robinson \& Wood}{Marsh, Robinson \& Wood}{1994}]{marsh1994}
Marsh T., Robinson E., Wood J.,  1994, MNRAS, 266, 137

\bibitem[\protect\citeauthoryear{Patterson}{Patterson}{2005}]{patterson2005}
Patterson J.,  2005, PASP, 117, 1204

\bibitem[\protect\citeauthoryear{Ritter \& Kolb}{Ritter \& Kolb}{2003}]{ritter2003}
 Ritter H., Kolb U., 2003, VizieR Online Data Catalog, 5113

\bibitem[\protect\citeauthoryear{Savoury et~al}{Savoury et~al.}{2011}]{savoury2011}
Savoury C. et~al., 2011, MNRAS, 415, 2025

\bibitem[\protect\citeauthoryear{Smak}{Smak}{1979}]{smak1979}
Smak J., 1979, ACTA, 29, 309

\bibitem[\protect\citeauthoryear{Smith, Dhillon \& Marsh}{Smith et~al.}{1998}]{smith1998}
Smith D.,  Dhillon V.,    Marsh T.,  1998, MNRAS, 296, 465

\bibitem[\protect\citeauthoryear{van Spaandonk, Steeghs, Marsh \& Torres}{van Spaandonk et~al.}{2010}]{spaandonk2010}
van Spaandonk L., Steeghs D., Marsh T., Torres M., 2010, MNRAS, 401, 1857

\bibitem[\protect\citeauthoryear{Tappert, Augusteijn \& Maza}{Tappert et~al.}{2004}]{tappert2004}
Tappert C.,  Augusteijn T.,    Maza J.,  2004, MNRAS, 354, 321

\bibitem[\protect\citeauthoryear{Thoroughgood, Dhillon, Littlefair, Marsh \& Smith}{Thoroughgood et~al.}{2001}]{thoroughgood2001}
Thoroughgood T.,  Dhillon V.,  Littlefair S.,  Marsh T.,    Smith D.,  2001,
  MNRAS, 327, 1323

\bibitem[\protect\citeauthoryear{Thoroughgood, Dhillon, Watson, Buckley,
  Steeghs \& Stevenson}{Thoroughgood et~al.}{2004}]{thoroughgood2004}
Thoroughgood T.,  Dhillon V.,  Watson C.,  Buckley D.,  Steeghs D.,
  Stevenson M.,  2004, MNRAS, 353, 1135

\bibitem[\protect\citeauthoryear{Tulloch, Rodr{\'{\i}}guez-Gil \& Dhillon}{Tulloch et~al.}{2009}]{tulloch2009}
Tulloch S.,  Rodr{\'{\i}}guez-Gil P.,    Dhillon V.,  2009, MNRAS, 397, L82

\bibitem[\protect\citeauthoryear{Wade \& Horne}{Wade \& Horne}{1988}]{wade1988}
Wade R.,  Horne K.,  1988, ApJ, 324, 411

\bibitem[\protect\citeauthoryear{Watson, Dhillon, Rutten \& Schwope}{Watson et~al.}{2003}]{watson2003}
Watson C., Dhillon V., Rutten R., Scwope A., 2003, MNRAS, 219, 629

\bibitem[\protect\citeauthoryear{Wood \& Horne}{Wood \& Horne}{1990}]{wood1990}
Wood J.,  Horne K., 1990, MNRAS, 242, 606

\bibitem[\protect\citeauthoryear{Wood, Irwin \& Pringle}{Wood et~al.}{1985}]{wood1985}
Wood J.,  Irwin M.,    Pringle J.,  1985, MNRAS, 214, 475

\bibitem[\protect\citeauthoryear{Wood, Horne, Berriman, Wade, O'Donoghue \& Warner}{Wood et~al.}{1986}]{wood1986}
Wood J.,  Horne K.,  Berriman G.,  Wade R.,  O'Donoghue D.,    Warner B.,
  1986, MNRAS, 219, 629
\end{thebibliography}
\end{document}